\documentclass[reprint,aps,prd,superscriptaddress,showpacs,preprintnumbers,amsmath,amssymb,floatfix]{revtex4-2} 

\usepackage[euler]{textgreek}
\usepackage[mathlines]{lineno}
\usepackage{graphicx,subfigure}
\usepackage{epstopdf}
\usepackage{dcolumn}
\usepackage{bm}
\usepackage{color}
\usepackage{upgreek}
\usepackage{afterpage}
\usepackage{placeins}
\usepackage{tikz, pgfplots}
\usepackage[version=4]{mhchem} 
\usepackage[colorlinks=true]{hyperref}
\usepackage[user,titleref]{zref}
\usepackage{tabularx}

\pgfplotsset{compat=1.18}

\newcommand{\keVee}{keV$_{\text{ee}}$} 
\newcommand{\MeVee}{MeV$_{\text{ee}}$} 

\newcommand{\geant}{\texttt{Geant4}}

\newcommand{\hmone}{\texttt{highmass1}}
\newcommand{\hmtwo}{\texttt{highmass2}}
\newcommand{\hmd}{\texttt{highmass}}
\newcommand{\lm}{\texttt{lowmass}}

\newcommand{\Efit}{\ensuremath{E_{\text{fit}}}}
\newcommand{\sigfit}{\ensuremath{\sigma_{\text{fit}}}}

\begin{document}


\title{Novel coincidence detection technique for precision measurement of\\ neutron capture–induced nuclear recoils}

\author{A.J.~Biffl}\email{Corresponding author: ajbiffl3@tamu.edu} \affiliation{Department of Physics and Astronomy, and the Mitchell Institute for Fundamental Physics and Astronomy, Texas A\&M University, College Station, TX 77843, USA}
\author{Gerardo~D~Gonzalez} \affiliation{Department of Physics and Astronomy, and the Mitchell Institute for Fundamental Physics and Astronomy, Texas A\&M University, College Station, TX 77843, USA}
\author{A.N.~Villano} \affiliation{Department of Physics, University of Colorado Denver, Denver, Colorado 80217, USA}
\author{N.~Mirabolfathi} \affiliation{Department of Physics and Astronomy, and the Mitchell Institute for Fundamental Physics and Astronomy, Texas A\&M University, College Station, TX 77843, USA}

\smallskip
\date{\today}

\noaffiliation


\smallskip

\begin{abstract}
The lattice dynamics following particle interactions remain not fully understood, and effects from nuclear-recoil interactions in conventional solid-state detectors — such as defect formation — can hinder accurate event-energy interpretation.
Neutron capture $(n,\gamma)$ can produce detectable, precise-energy nuclear recoils for detector calibration and solid-state physics studies. This paper proposes a first-of-its-kind experiment to measure the twin products of neutron capture -- the postcapture deexcitation gamma and postcapture nuclear recoil.
Simulations show that with a 1\,mCi californium neutron source, a positive measurement with existing detector technologies is possible in 26.1\,gram-days of exposure. 
\end{abstract}


\maketitle

%
%
%
%
%
%
%

%
%
%
%
%
%
%

\section{\label{sec:intro}Introduction}

Crystal-based particle detectors 
have applications for dark matter searches~\cite{10.1140/epjc/s10052-016-3877-3, 10.1103/PhysRevLett.121.051301,  10.1103/PhysRevLett.125.241803, PhysRevD.106.062004, 10.1103/physrevlett.121.061803, 10.1016/j.nuclphysb.2024.116465}, 
neutrino science~\cite{10.1103/PhysRevD.91.072001, 10.1140/epjc/s10052-021-09038-3, 10.1103/PhysRevD.104.072003, 10.1016/j.nima.2021.165342, 10.1103/PhysRevD.96.022009, 10.1088/1748-0221/10/12/P12011, 10.1007/s12648-017-1004-4}, and nuclear energy~\cite{10.1016/j.net.2016.02.001}. 
In these detectors, especially ones that measure nuclear recoils (NRs), specific knowledge of the dynamics of atoms in the crystal lattice is helpful but is locked behind crystallographic processes. Details about particle motions and energies are obscured, making it difficult to interpret collected data. It is therefore helpful to study events where the energy is known beforehand, to aid in reconstructing the dynamics of the event and to calibrate these detectors.  

There is still much unknown about the interactions between a recoiling atom and its surrounding crystal lattice during a NR. Formation of crystal defects, direction-dependent effects on recoil energies (e.g., channeling~\cite{1475-7516/2010/03/029}), and ionization yield are difficult to measure without detailed information about the NRs in question. 

Crystal defects can warp the spectrum of NR events from DM (dark matter) or CENNS (coherent elastic neutrino-nucleus scatter)~\cite{PhysRevD.111.085021} and uncertainty about the energy, time scale, and evolution of defects can hinder experiments and may even be a major source of backgrounds~\cite{, PhysRevD.106.083009, 10.1063/5.0247343}.
Defect properties have been studied extensively in silicon, with work focusing on defects caused by implanted silicon ions~\cite{10.1063/1.1763242} or low-energy helium~\cite{10.1063/1.5046096}, defect annealing after electron irradiation~\cite{10.1007/s11664-018-6286-6}, comparison between electron and proton irradiation~\cite{10.1134/S1063782612040069}, effects of an external electric field~\cite{10.1016/S0168-583X(02)01817-7,10.1134/1.1187794},  and defect concentration after hydrogen implantation~\cite{10.1016/j.nimb.2013.09.045}. However, in all these cases the defect properties had to be characterized by measuring bulk material properties (c.f.,~\cite{10.1038/s43246-023-00379-y, 10.1063/5.0077299}), rather than directly studying the dynamics of defect formation, evolution, and annealing. 

Ionization yield, the fraction of NR energy given to ionizing electrons, is an extremely important parameter for ionization-based detectors. 
The most common model for the ionization yield is the Lindhard model~\cite{osti_4701226}, 
though recent measurements have shown disagreement with the Lindhard model in Si at low energies ($\lesssim$ keV$_\text{ee}$)~\cite{10.1103/PhysRevD.102.063026, 10.1103/PhysRevD.94.082007, 10.1088/1748-0221/12/06/P06014, 10.1103/PhysRevD.105.083014, PhysRevLett.131.091801}. 
Several modifications have been tried to make up these differences~\cite{PhysRevD.91.083509, 10.1103/PhysRevD.105.122002}, though they are ultimately limited by lack of measurement below around 100\,eV. Uncertainty in the yield introduces large uncertainties in measured energies and calibrations. This can cause large uncertainties in thresholds and event rates, and thus in reported dark matter limits (c.f., \cite{10.1103/PhysRevD.99.062001} \S{}IIA and \S{}VIIIB, and \cite{10.1103/PhysRevD.97.022002} \S{}IIA and \S{}VA\&B).

Thermal neutron capture $(n,\gamma)$ has been proposed as a source of exact-energy NRs for detector calibration~\cite{10.1103/PhysRevD.107.076026, 10.1007/s10909-022-02816-7, PhysRevD.106.032007}; it may also serve as a door into direct measurement of defect formation, ionization yield, and other crystal effects in NRs. The excited-state postcapture nucleus emits one or more MeV-scale deexcitation gammas and recoils at its position in the crystal lattice. If this deexcitation happens in a single gamma (a ``straight-to-ground" event),
the NR will have an exactly known energy, which can be used for detector calibrations and other studies. 
Capture-induced NRs have previously been measured in silicon~\cite{10.1103/PhysRevD.105.083014}, and the straight-to-ground NR peak from capture on $^{182}$W in $\text{CaWO}_4$ has been recently measured by the CRAB Collaboration~\cite{10.1103/PhysRevLett.130.211802, 10.1007/s10909-024-03252-5}, but the straight-to-ground peak has never been measured in other common detector materials like germanium or silicon.

In this paper we propose a new experiment to measure the capture-induced NR signature in silicon and the postcapture deexcitation gammas in coincidence using a gram-scale silicon phonon detector and two kg-scale germanium ionization detectors. 
This will be the first measurement of the NR peak from capture-induced recoils in silicon, and will be the first time a capture-induced NR is measured in coincidence with the emitted deexcitation gamma. 
The structure of the paper is as follows: Section \ref{sec:cap_spec} describes the signature of neutron capture in silicon in detail. 
In Section \ref{sec:setup} the proposed experimental configuration is described. 
Section \ref{sec:simulation} details simulations carried out to estimate the expected signal rate, background rate, and detector requirements. Discussion and concluding remarks are given in Section \ref{sec:conclusions}.

\section{\label{sec:cap_spec}Neutron Capture}


During radiative thermal neutron capture $(n,\gamma)$, a nucleus absorbs a low-energy ($O(10^{-2}\text{\,eV})$) neutron. The neutron enters the nuclear system at the neutron separation energy, $S_n$ ($\sim$\,MeV), and very quickly ($\sim$\,fs) after the capture the nucleus deexcites to its ground state, emitting a cascade of one or more gammas. 
Each gamma emission causes a corresponding momentum change of the nucleus, though the cascade is so fast that all of the NR energy comes in a single pulse for all practical detectors.


Each individual gamma emission in the deexcitation follows two-body kinematics, but unless the nucleus's momentum is known at all moments in the intermediate phases of a multi-gamma cascade, reconstructing the exact recoil energy and direction is impossible. However, in the case of a single-gamma cascade we \emph{can} reconstruct the exact NR energy and direction, assuming the nucleus is at rest before the cascade. 
In this case, the postcapture nucleus falls directly to the ground state, and the sum of gamma energy and recoil energy is equal to the neutron separation energy. 
The direction of the NR is opposite the emitted gamma direction. For neutron capture on a nucleus with rest mass $M$ and neutron separation energy $S_n$, the gamma energy $E_\gamma$ is:

\begin{equation}
    E_\gamma = -M + \sqrt{M^2 + 2MS_n}\simeq S_n - \frac{S_n^2}{2M}
\end{equation}

\noindent and the kinetic energy of the recoiling nucleus $E_{nr}$ is:

\begin{equation}
    E_{nr} = \frac{E_\gamma^2}{2M} \simeq \frac{S_n^2}{2M}
\end{equation}

\noindent For $^{28}\text{Si}(n,\gamma)^{29}\text{Si}$, these energies $S_n$, $E_\gamma$, and $E_{nr}$ are given in Table~\ref{tab:capture-energies}. At finite temperature the energies are broadened by an amount $\Delta E\approx S_n\sqrt{T/M}$. At 20\,mK, this broadening is $\approx0.067$\,eV, or $10^{-6}\%$ of the gamma energy and $10^{-2}\%$ of the NR energy, and can safely be neglected. 
The gamma and NR deposits are thus almost perfectly monoenergetic after a straight-to-ground cascade, and their momentum directions are almost perfectly opposite each other. With an appropriate experimental setup, it should be possible to measure the characteristic peak of these straight-to-ground events. 
Other straight-to-ground transitions are possible to measure but occur at reduced frequency (natural Si contains 92.2\% $^{28}$Si, compared to 4.7\% $^{29}$Si and 3.1\% $^{30}$Si). $S_n$, $E_\gamma$, and $E_{nr}$ for capture on these other isotopes are also listed in Table~\ref{tab:capture-energies}.

\begin{table}[htbp]
    \centering
    \begin{ruledtabular}
    \begin{tabular}{cccc}
        Isotope & Si-29 & Si-30 & Si-31 \\
        \hline 
        $S_n$ & 8473.6  & 10609.2  & 6587.4  \\
        \hline
        $E_\gamma$ & 8472.27  & 10607.18  & 6586.64  \\
        \hline
        $E_{nr}$ & 1.32966  & 2.01487  & 0.75179  \\
    \end{tabular}
    \end{ruledtabular}
    \caption{Neutron separation energy $S_n$, postcapture gamma energy $E_\gamma$, and NR energy $E_{nr}$ for postcapture isotopes $^{29}\text{Si}$, $^{30}\text{Si}$, and $^{31}\text{Si}$. $S_n$ values from EGAF~\cite{10.1063/1.2187849} and isotope masses from NuDat\,3~\cite{nudat}. All energies in keV.}
    \label{tab:capture-energies}
\end{table}


\section{\label{sec:setup}Experimental Configuration}

To measure these events, we propose a three detector setup consisting of a gram-scale silicon phonon detector to measure the postcapture nuclear recoil and two large-mass (kg-scale) germanium ionization detectors to measure the emitted postcapture gammas. We refer to the Si phonon detector as \lm{} and the Ge gamma detectors as \hmone{} and \hmtwo{}. 
The \lm{} detector will use tungsten transition edge sensors (TESs) with aluminum fins photolithographically deposited on the silicon to read athermal phonon signals in the substrate. 
We have already demonstrated phonon energy resolution on the order of a few eV at threshold in this type of detector~\cite{10.1103/PhysRevD.104.032010, 10.1016/j.nima.2020.163757, 10.1016/j.nima.2021.165489}.
The NR energy resolution is strongly influenced by variation in phonon collection efficiency with event position, but multichannel arrangement of the TESs on the surface allows for reliable event position reconstruction
(see, e.g., \cite{10.1103/PhysRevD.97.022002} \S{}IIC and \cite{10.1103/PhysRevD.92.072003} \S{}IIIB), and we expect to be able to maintain sub-percent resolutions up to keV-scale energies. It is also possible to use NTL-assisted phonon readout to measure NR ionization energy in this detector~\cite{10.1016/0168-9002(90)91510-I, neganov1985colorimetric}.
Alternatively, CUORE has demonstrated detector resolutions on the order of 0.1\% for thermal phonons (i.e., measuring temperature change) at MeV scales in TeO$_2$ crystals~\cite{10.1140/epjc/s10052-014-2956-6}. Thermal phonon detectors have longer pulse times (affecting timing resolution) but are immune from position-dependent effects.

The germanium \hmd{} detectors will be kg-scale ionization detectors. Ionization allows much faster detector response than phonons 
($\sim1$\,$\mu$s rise time), meaning that even with larger backgrounds in these detectors than in \lm{}, pileup and coincidence identification will not be problems. Typical resolutions in this type of ionization detector at MeV energies are on the order of 0.1\% and generally scale with $\sqrt{E_\gamma}$ (c.f., \cite{10.1016/j.mex.2016.12.003, 10.1016/j.mex.2016.12.003, 10.1016/j.nima.2008.04.017, 10.1016/S0168-583X(03)01594-5} for operation at $\sim77$K. There is expected to be some degradation in resolution at sub-Kelvin temperatures but not substantial~\cite{phipps_phd}). 
The mean free path of 8\,MeV gammas in germanium is $\sim$ 6\,cm~\cite{XCOM_database}, which is the approximate length scale of a 0.6-kg germanium detector (see detector dimensions in Section~\ref{sec:simulation}).
For both detector types, the expected signal energies (8.47\,MeV ionization and 1.33\,keV NR) are well above typical detector thresholds.

\begin{figure}[htbp]
    \centering
    \includegraphics[width=0.7\linewidth]{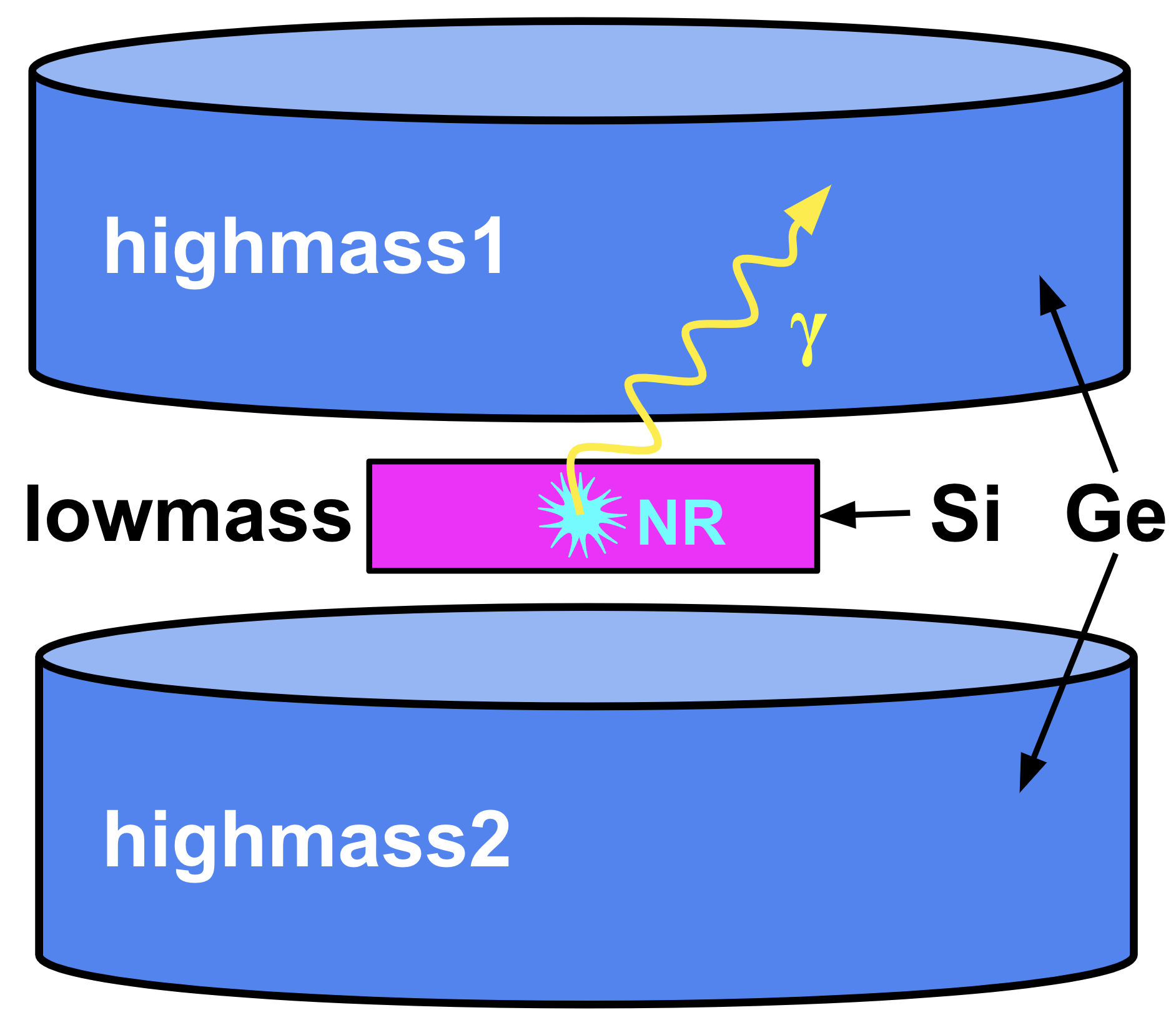}
    \caption{Schematic of three-detector stack including the \hmd{} germanium ionization detectors (blue) and central silicon phonon detector (magenta). The NR resulting from neutron capture will be detected in \lm{} and the postcapture deexcitation gamma will be detected in coincidence by one of the \hmd{} detectors.}
    \label{fig:detector-schematic}
\end{figure}

In the proposed experiment, \lm{} will be sandwiched between the two gamma detectors (see Figure~\ref{fig:detector-schematic}) to maximize the solid angle attenuated (around $95\%$ solid angle 
($3.79\pi$ steradians) for simulated dimensions, see Section~\ref{sec:simulation}). Even larger solid-angle coverage can also be achieved with an annular germanium detector~\cite{10.21468/SciPostPhysProc.12.017}. The detector assembly will be loaded into a dilution refrigerator for cryogenic operation ($\sim20$\,mK).

MeV-scale neutron flux will be provided by a 1\,mCi~(37\,MBq) $^{252}$Cf sample, an accessible neutron emitter~\cite{10.1016/S0969-8043} that is cheaper than alternatives like DD-generators or nuclear reactors, and provides larger flux than other radioactive sources such as AmBe. The Cf source will be placed outside of the refrigerator behind a two-component shield consisting of lead to block gammas from the Cf decay chain and high-density polyethylene (HDPE) to thermalize the neutrons. Despite the shielding, it is expected that the $^{252}$Cf will be the dominant source of backgrounds (any events other than neutron captures which deposit energy in the detectors). The main environmental backgrounds are cosmic rays and environmental radioactivity, which will be small ($\sim$100\,Hz in the \hmd{} detectors) compared to the source-coincident backgrounds and so are ignored.

\section{\label{sec:simulation}Simulation}

\begin{figure}[htbp]
    \centering
    \includegraphics[width=0.9\linewidth]{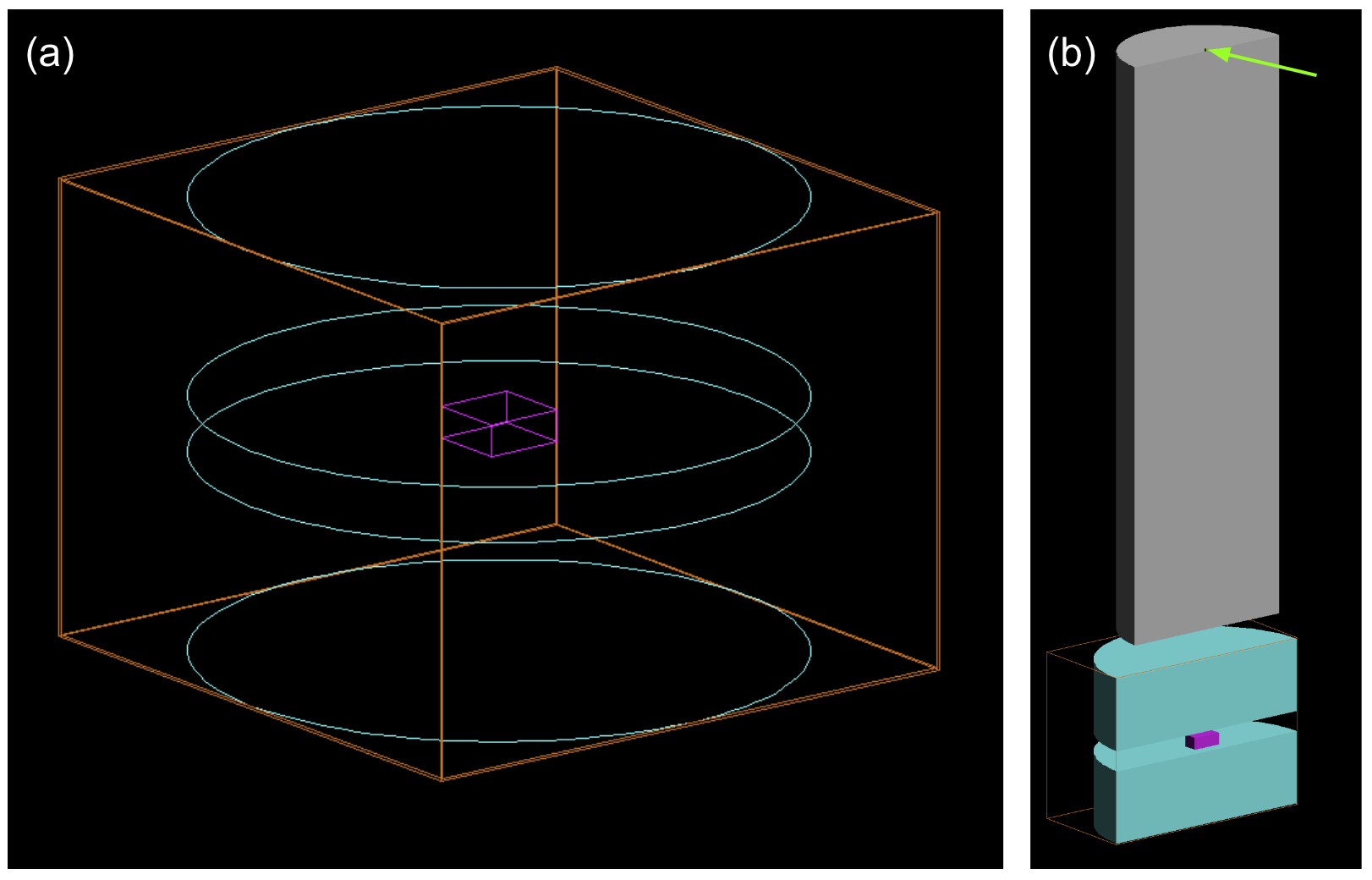}
    \caption{Visualized geometry of \geant{} simulation (see text). \textbf{(a)} detector stack consisting of twin germanium gamma detectors (turquoise), silicon phonon detector (magenta), surrounded by copper box (orange). 
    \textbf{(b)} sliced view of detectors (turquoise/magenta) and lead shielding (gray) with source cube barely visible on top of lead shield (green arrow). In the actual experiment the source and shielding will likely be placed directly below the fridge can, rather than above as shown.
    }
    \label{fig:simulation_geometry}
\end{figure}

A minimal experimental setup was modeled in \geant{} version 4.11.3~\cite{1610988, AGOSTINELLI2003250} to estimate the postcapture gamma signal rate and the source-coincident backgrounds. The simulation geometry is shown in Figure~\ref{fig:simulation_geometry}, and consists of the detector stack, lead shielding, source cube, and world volume. 
The \hmd{} detectors are 7.5\,cm diameter, 2.5\,cm thick germanium cylinders (588\,g) and the \lm{} detector is a 1\,cm$\times$1\,cm$\times$4\,mm silicon box (0.932\,g). The detectors are surrounded by a thin (0.2\,mm) copper box. The empty space inside the copper box is vacuum. There is a 1.5\, mm gap between the silicon and germanium detectors to give space for instrumentation and readout cables. The lead shield is a 20\,cm thick, 2.5\,cm diameter cylinder spaced 1\,cm above the detector box. The source cube is a 1\,mm californium cube centered on the top face of the lead shield. The world box is composed of HDPE, and the 1\,cm gap between the lead shielding and the detector box serves to emulate the HDPE component of the full shield.

The emission spectrum of a 1\,mCi (37\,MBq) californium source was simulated by generating $^{252}$Cf nuclei inside the source cube volume. 
The ``\verb+Shielding_HPT+" physics list was used, which on top of the standard electromagnetic and hadronic interactions employs the NeutronHP (High-Precision) evaluated data libraries for neutron interaction cross sections and lifetimes~\cite{mendoza_cano-ott_iaea}, as well as the \texttt{NeutronHPThermalScattering} process for low energy ($<4$\,eV) neutrons. This physics list models the radioactive decay and subsequent decay products of the $^{252}$Cf nuclei. 
Note that the hadronic radioactive decay time threshold (the maximum time of decay for products to be tracked) had to be increased from the default value 
to see all the Cf decays, though extremely late-arriving events ($t>100$\,yr) were thrown out to avoid overcounting the decays of $^{248}$Cm, the long-lived ($T_{1/2}\sim10^5$\,yr) daughter isotope of $\alpha$-decay from $^{252}$Cf. 
%
Lastly, the \texttt{PhotonEvaporation} model was enabled for nuclear deexcitations (e.g., postcapture cascades) 
to conserve energy in postcapture cascades.

\subsection{Shield optimization}

\begin{figure}[htbp]
    \centering
    \includegraphics[width=\linewidth]{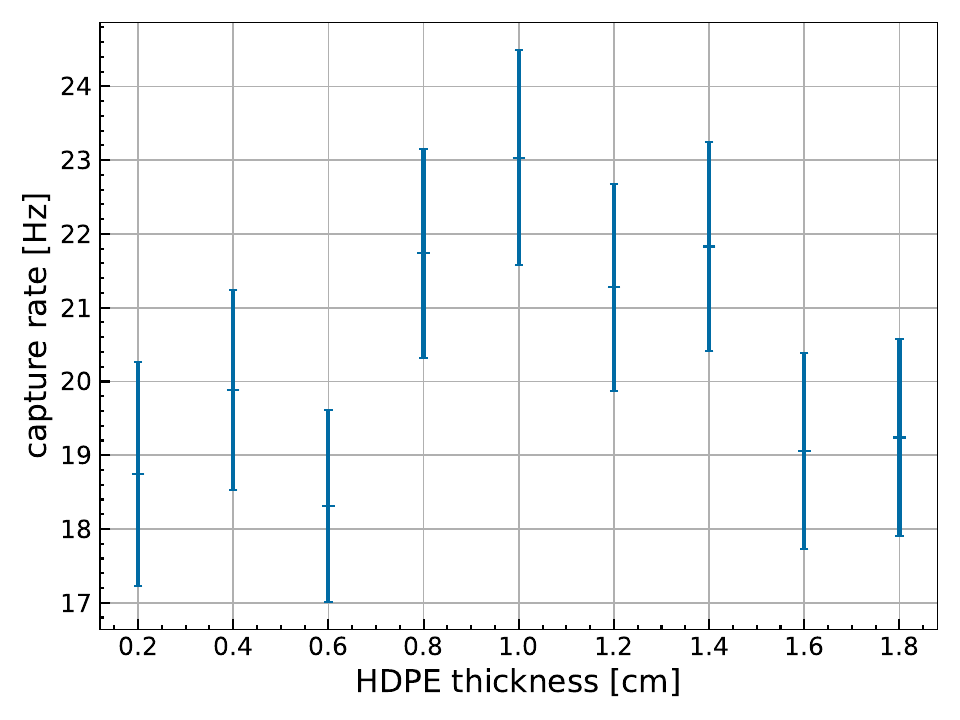}
    \caption{Neutron capture rate in \lm{} vs. HDPE thickness for simulations with 5\,cm total (lead+HDPE) between source and detector box. The lead has negligible effect on the neutrons. The ideal HDPE thickness is about 1\,cm.}
    \label{fig:poly_tuning}
\end{figure}

The final shielding parameters (20\,cm Pb and 1\,cm HDPE) were tuned individually during dedicated simulations. The HDPE thickness was tuned to maximize the neutron capture rate in the low mass detector given a fixed total distance between source and detector box. This distance was set to 5\,cm total (lead+HDPE) to give large statistics in the capture rate. As shown in Figure~\ref{fig:poly_tuning}, the maximum capture rate was found with around 1\,cm of HDPE. We expect this to be true for larger separation between source and detector as well, as the lead has little effect on the neutrons.

\begin{figure}[htbp]
    \centering
    \includegraphics[width=\linewidth]{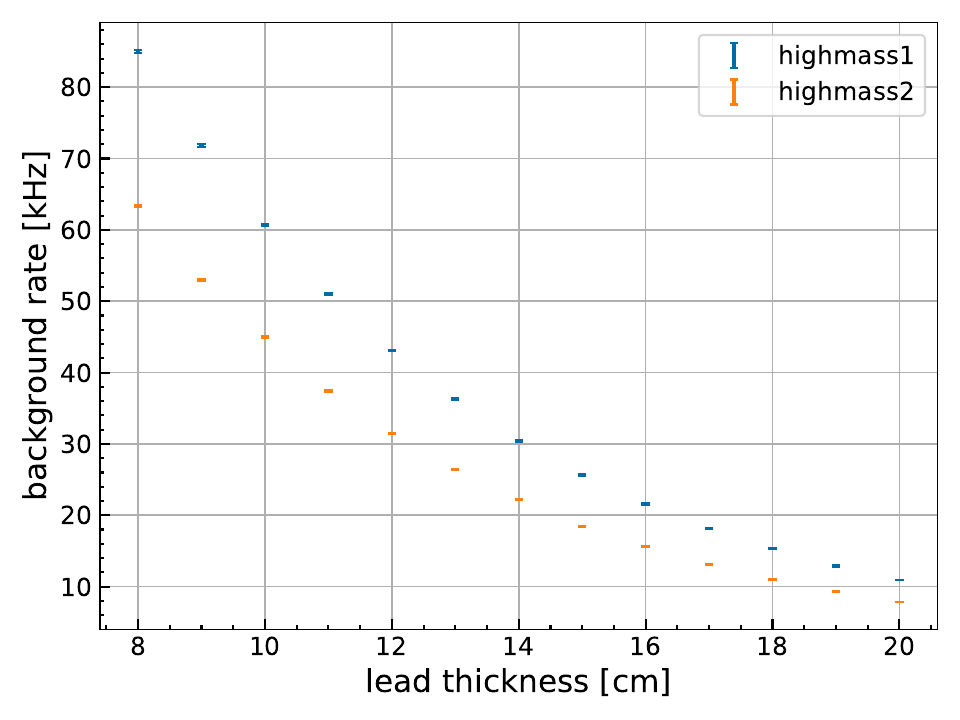}
    \caption{Background rate above 100\,\keVee{} in \hmd{} detectors for different thickness of lead. With 20\,cm of lead, the target background of $\sim$10\,kHz is achieved.}
    \label{fig:lead_tuning}
\end{figure}

Given 1\,cm HDPE, the lead thickness was tuned so that the background rate in \hmone{} is low enough for consistent coincidence identification with events in \lm{}. The rise time in the \lm{} detector is approximately 100\,$\mu$s, so to identify events in coincidence with a given pulse in \lm{}, pulses have to be arriving in \hmone{} further than $\sim100$\,$\mu$s apart, giving a target background rate of $\sim10$\,kHz. Events of interest in \hmone{} are $\sim$MeV energy, so we only count events $\geq100$\,\keVee{} in this rate (electron-equivalent or ionization energy calculated for simulated events with the Lindhard model~\cite{osti_4701226}). Figure~\ref{fig:lead_tuning} shows the background rate in \hmone{} above 100\,\keVee{} for different lead thicknesses. The target rate of 10\,kHz is approximately reached with 20\,cm of lead.

\subsection{Source/shield simulations}

With the shield parameters finalized, the simulation was split into two phases. In the first phase, the emission spectrum of the californium source was simulated propagating through the shield to the detector box. 
Hits (energy deposits) were recorded in the three sensitive volumes -- \hmone{}, the germanium detector nearest the source, \hmtwo{}, the germanium detector furthest from the source, and \lm{}, the central silicon phonon detector. For an exposure of 540 seconds ($2\times10^{10}$ primaries), there were 495 neutron captures in \lm{}, meaning a capture rate of roughly ($0.92\pm0.04$)\,Hz. 

\begin{figure}[htbp]
    \centering
    \includegraphics[width=\linewidth]{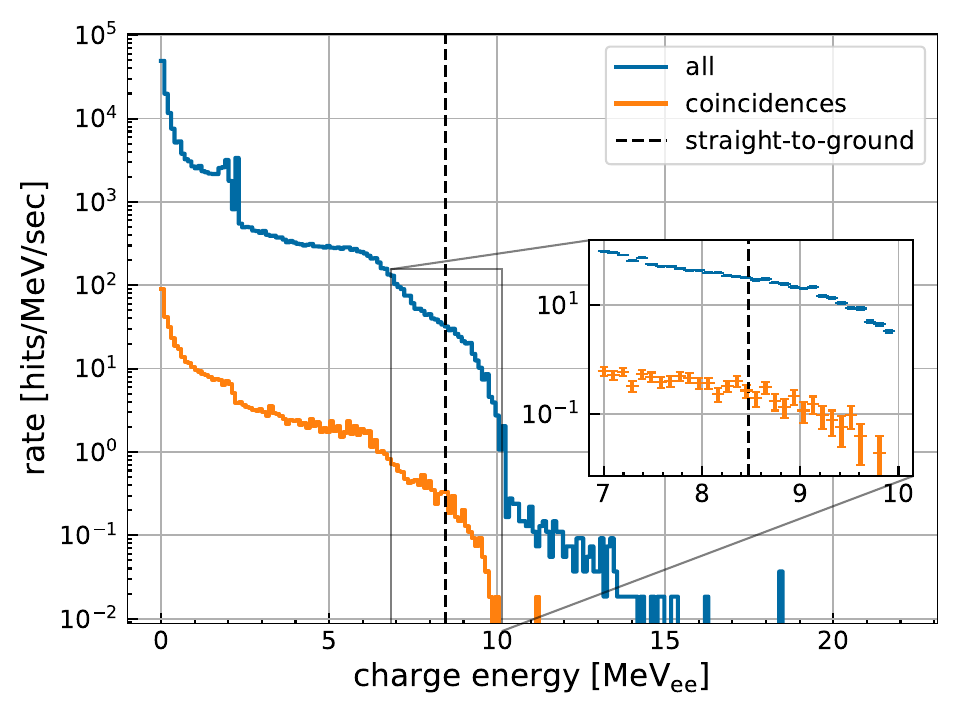}
    \caption{Background spectrum in \hmone{} detector after different rounds of cuts. \textbf{(blue)} all events, 10.92\,kHz; \textbf{(orange)} coincidence events, 38.81\,Hz (rates for energies above 100\,keV). \textbf{(inset)} Energy cut window. Total rate is 109.4\,Hz; coincidence rate is 0.814\,Hz.}
    \label{fig:sim-highmass1-background}
\end{figure}

The background spectrum in \hmone{} is shown in Figure \ref{fig:sim-highmass1-background}. The upper curve shows all events that deposited energy in \hmone{}, and the ``coincidence" events are defined to be any events where energy is deposited in both the \lm{} detector and either one of the \hmd{} detectors. 
The inset shows the spectrum of events in a wide energy cut region surrounding the expected straight-to-ground signal, between 7\,\MeVee{} and 10\,\MeVee{}.
The total event rates and rates in the energy cut window are given in Table~\ref{tab:sim-background-rates}. The table also shows event rates in \lm{}.

\begin{table}[htbp]
    \centering
    \begin{ruledtabular}
    \begin{tabular}{ccc}
        \multicolumn{3}{c}{\hmone{} rates (Hz)}\\
        \hline
         energy range & $>100$\,\keVee{} & cut window \\
         \hline
         all & $10,924.6\pm4.5$ & $109.44\pm0.45$ \\
        \hline
        coincidences & $38.81\pm0.27$ & $0.814\pm0.039$ \\
    \end{tabular}
    \begin{tabular}{cccc}
        \multicolumn{4}{c}{\lm{} rates (Hz)}\\
        \hline
        energy range & all energies & $<3$ keV  & near peak \\
         \hline
        all & $54.49\pm0.32$ & $4.209\pm0.088$ & $0.461\pm0.029$ \\
        \hline
        coincidences & $50.89\pm0.31$ & $3.689\pm0.083$ & $0.411\pm0.028$ \\
        \hline
        energy cut & $1.210\pm0.047$ & $0.070\pm0.011$ & $0.019\pm0.006$ \\
    \end{tabular}
    \end{ruledtabular}
    \caption{Event rates in \hmone{} above $100$\,\keVee{} and in the cut window $7\text{\,\MeVee{}}\leq E\leq 10\text{\,\MeVee{}}$, as well as event rates in \lm{} for all energies (the highest-energy events were a little over 6\,MeV), as well as in the lower-energy region of interest $<3$\,keV and in a 500\,eV window near the $^{29}$Si straight-to-ground NR peak, $1.08\text{\,keV}\leq E\leq 1.58\text{\,keV}$. }
    \label{tab:sim-background-rates}
\end{table}

\begin{figure}
    \centering
    \includegraphics[width=\linewidth]{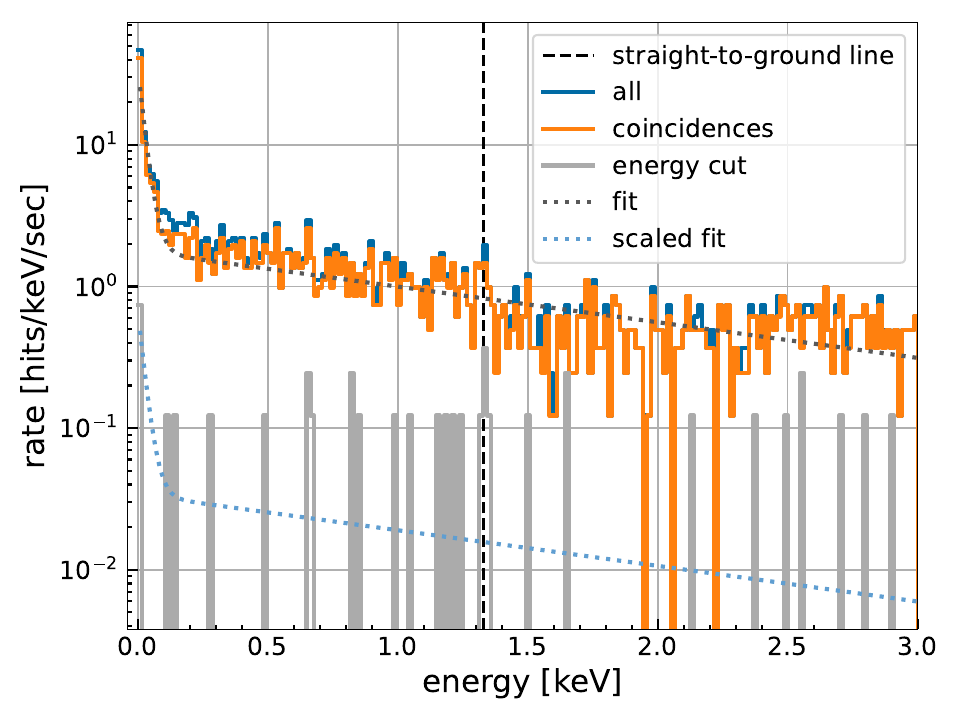}
    \caption{Background spectra in \lm{} below 3\,keV. The total rate below 3\,keV is 4.209\,Hz, while the coincidence rate is 3.689\,Hz. After energy cuts the rate is 70\,mHz. Plot also shows double-exponential spectrum from equation (\ref{eqn:logdoubleexp}) fitted to the coincidence data, and that fit rescaled to the rate after energy cuts.}
    \label{fig:sim-lm-spectra}
\end{figure}

Figure \ref{fig:sim-lm-spectra} shows the background spectra in \lm{} below 3\,keV, including events remaining after energy cuts in the \hmd{} detectors between 7\,\MeVee{} and 10\,\MeVee{} (the cut window shown in the inset of Figure~\ref{fig:sim-highmass1-background}). 
The log of the ``coincidence" spectrum in Figure~\ref{fig:sim-lm-spectra} was fitted to a double-exponential of the form:

\begin{equation}
\label{eqn:logdoubleexp}
    \log H  = \log \left(Ae^{-E/a} + Be^{-E/b}\right)
\end{equation}

\noindent for the spectrum height $H$. The best-fit values for the parameters $A$, $a$, $B$, and $b$ are given in Table~\ref{tab:doubleexp-fits}. The resulting fit is also plotted in Figure~\ref{fig:sim-lm-spectra}, as well as the fit rescaled to the event rate after energy cuts (0.07\,Hz from 3.689\,Hz).

\begin{table}[htbp]
    \centering
    \begin{ruledtabular}
    \begin{tabular}{ccccc}
        Parameter & $A$ & $a$ & $B$ & $b$ \\
        \hline
        Fit value & 33.66 & 0.0246 & 1.78 & 1.73\\
        \hline
        Scaled value& 0.641 & 0.0246 & 0.0339 & 1.73\\
    \end{tabular}
    \end{ruledtabular}
    \caption{Best fit parameters for a double exponential (\ref{eqn:logdoubleexp}) fit to the coincidence spectrum in \lm{} below 3\,keV (Figure~\ref{fig:sim-lm-spectra}), and the parameters for the fit scaled to the rate after energy cuts. Here $A$ and $B$ are in units of events/keV/sec, $a$ and $b$ are in units of keV.}
    \label{tab:doubleexp-fits}
\end{table}

We will also focus on a ``near peak" region in the \lm{} detector in a 500\,eV window around the straight-to-ground NR peak 
(the last column in Table~\ref{tab:sim-background-rates}), $1.08\text{\,keV}\leq E\leq 1.58\text{\,keV}$, 
as the region of interest for searching for the straight-to-ground signal. 
The total background rate in this window is 0.461\,Hz, and the coincidence rate is 0.411\,Hz (compared to 0.414\,Hz predicted by the fit from table~\ref{tab:doubleexp-fits}). 
After energy cuts, there are only ten events present in this window ($\sim18.5$\,mHz, compared to 7.9\,mHz from the scaled fit).

\subsection{Postcapture gamma simulations}

\begin{figure}[htbp]
    \centering
    \includegraphics[width=\linewidth]{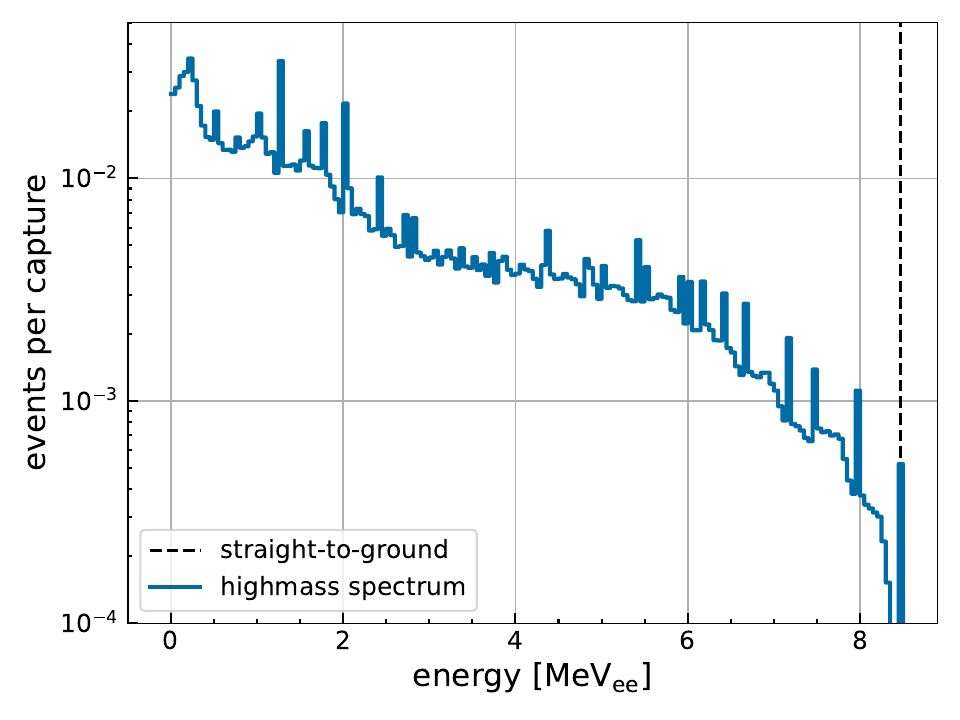}
    \caption{Spectrum of events in the \hmd{} detectors after $^{29}$Si neutron capture in \lm{}, normalized to the number of captures. The straight-to-ground gamma is fully collected in $(0.05199\pm0.00023)\%$ of captures.}
    \label{fig:sim-capture-gamma-spectrum}
\end{figure}

Because of the extremely low capture rate ($495$ captures over $2\times10^{10}$ primaries),
the rate of postcapture gamma detections was estimated in a separate phase of simulations. In this second phase, postcapture nuclei were sampled directly in the \lm{} detector by generating $^{29}$Si nuclei in an excited state with energy equal to the neutron separation energy, $S_n=8473.6$\,keV~\cite{10.1063/1.2187849}. 
Note this is not the same procedure that \geant{} uses to simulate neutron captures, but the resulting gamma spectra were nearly identical to the built-in \texttt{nCapture} process and there were significant gains in the efficiency of the simulations over other methods for generating the neutron captures directly. Figure~\ref{fig:sim-capture-gamma-spectrum} shows the spectrum of events in the \hmd{} detectors (sum of \hmone{} and \hmtwo{}) caused by deexcitation of the $^{29}$Si nuclei. In $0.052\%$ of captures, the full energy of the straight-to-ground gamma is detected. That gives a total signal rate of ($0.476\pm0.023$)\,mHz, or 41~events/day (compared to $\sim20$\,mHz backgrounds after energy cuts).

\subsection{\label{sec:analysis}Detector resolution simulations}

The simulated signal rate of ($0.476\pm0.023$)\,mHz 
was used to estimate the required detector resolution. To make these estimates, and because of the large uncertainties in the after-energy-cuts background rate, a ``worst-case" background event rate in the near-peak window was used at ten times the fitted rate ($78.8$\,mHz, about four times the simulated rate of 18.5\,mHz). Four-week exposures, containing 1,151 signal events and 190,648 background events each, were generated for a range of detector resolutions. 

\begin{figure}[htbp]
    \centering
    \includegraphics[width=\linewidth]{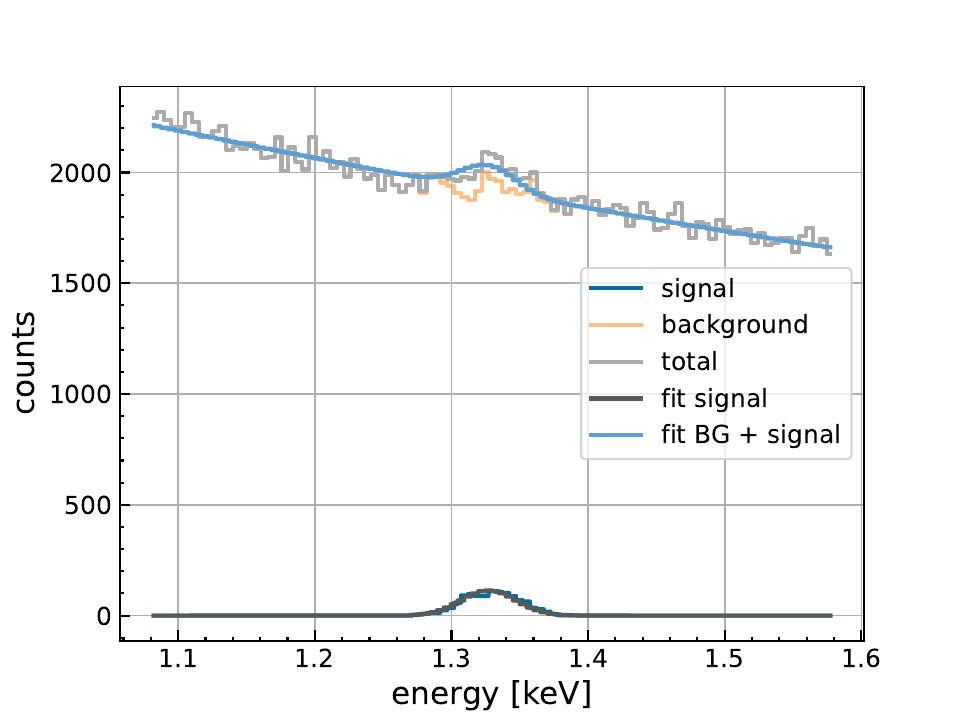}
    \caption{An example simulated four-week exposure in \lm{} with a resolution of 20\,eV, consisting of 1,151 signal events (dark blue) and 190,648 background events (orange). The total spectrum (light grey) is fitted (light blue) by maximizing a binned likelihood function. The extracted signal spectrum from the fit (dark grey) is pictured as well. The pictured 
    sigma value is $7.6\sigma$. The fitted signal mean was 1.335\,keV and the fitted width was 16.68\,eV.
    }
    \label{fig:sim-4wk-example}
\end{figure}

Signal events were modeled as a gaussian with the detector ``resolution" being the gaussian's width parameter. 
Background events were sampled from the scaled double-exponential fit from Table~\ref{tab:doubleexp-fits} between 1.08 and 1.58\,keV. 
An example simulated four-week exposure is shown in Figure~\ref{fig:sim-4wk-example} for a resolution of 20\,eV (comparable to our expected sensor resolution of $\sim10$\,eV). For each simulated four-week exposure, the deviation of the total spectrum (191,799 events) from a background-only spectrum with the parameters from Table~\ref{tab:doubleexp-fits} (and normalized to the correct total number of events) was characterized with a likelihood ratio test. 
Spectra composed of background and a gaussian peak are fitted to the generated spectra by maximizing the value of a binned likelihood function:

\begin{equation}
\label{eqn:likelihood}
    \mathcal{L} = e^{-(n_b+n_s)} \prod_{i}\frac{1}{C_i!} \big(n_b P_{b,i} + n_s P_{s,i}\big)^{C_i}
\end{equation}

\noindent where the index $i$ runs over each bin ($N_{bins}=100$), $n_b$ and $n_s$ are the total number of background and signal events respectively, $C_i$ is the count of events in each bin, and the $P_i$ are the probability mass of the background and signal spectra in each bin. 
We leave $P_{b,i}$ the background spectrum constant but allow the mean \Efit{} and width \sigfit{} of the gaussian signal peak to vary, $P_{s,i} = P_{s,i}(\Efit{},\sigfit{})$. The likelihood (\ref{eqn:likelihood}) is then maximized over four parameters $n_b$, $n_s$, $\Efit{}$, and \sigfit{}. 
The maximized likelihood value $\mathcal{L}_{max}$ is compared to $\mathcal{L}_b$, the likelihood of only the background spectrum  ($n_b=N_{total}=\sum_i C_i$ and  $n_s=0$). 
As per Wilks' theorem~\cite{Wilks:1938dza}, in the absence of a signal the quantity

\begin{equation}
    q = 2\ln\left(\frac{\mathcal{L}_{max}}{\mathcal{L}_b}\right)
\end{equation}

\noindent is distributed as a chi-squared variate with one degree of freedom and $\sqrt{q}$ is a standard normal variate. $\sqrt{q}$ is thus reported as a ``sigma" value giving the significance of a signal measurement~\cite{10.1140/epjc/s10052-011-1554-0} (see, e.g.,~\cite{10.1088/1742-6596/1342/1/012132}). A value of 5 denotes a typical ``$5\sigma$" significance.

\begin{figure}[htbp]
    \centering
    \includegraphics[width=\linewidth]{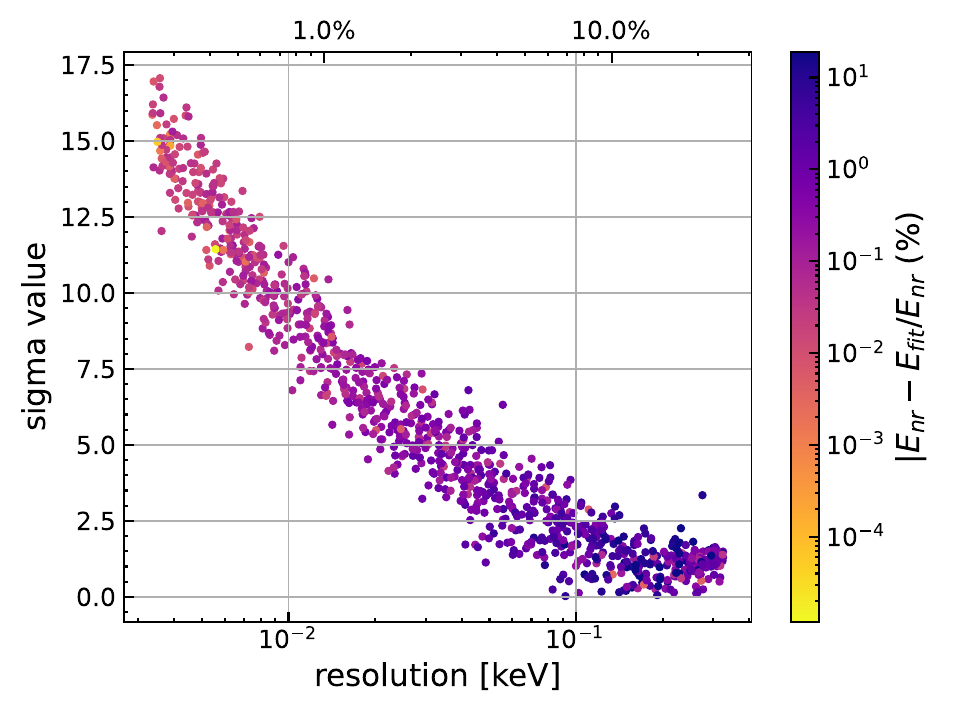}
    \caption{Sigma values for simulated four-week exposures in \lm{} with color giving relative error in the fitted value of $E_{nr}$, the signal peak position. The sigma values cross 5$\sigma$ below about 20\,eV resolution, or 1\%.}
    \label{fig:sim-4wk-colored-chi2}
\end{figure}

Figure~\ref{fig:sim-4wk-colored-chi2} shows sigma-values for 1,000 simulated four-week exposures and the error in the fitted peak positions from the true energy, $E_{nr}$, with randomly-generated detector resolutions. The simulated exposures reach $5\sigma$ significance below around 20\,eV resolution (1\%) in \lm{}.  This is well within the range of achievable resolutions for the detectors described in Section~\ref{sec:setup}.

\begin{figure}[htbp]
    \centering
    \includegraphics[width=\linewidth]{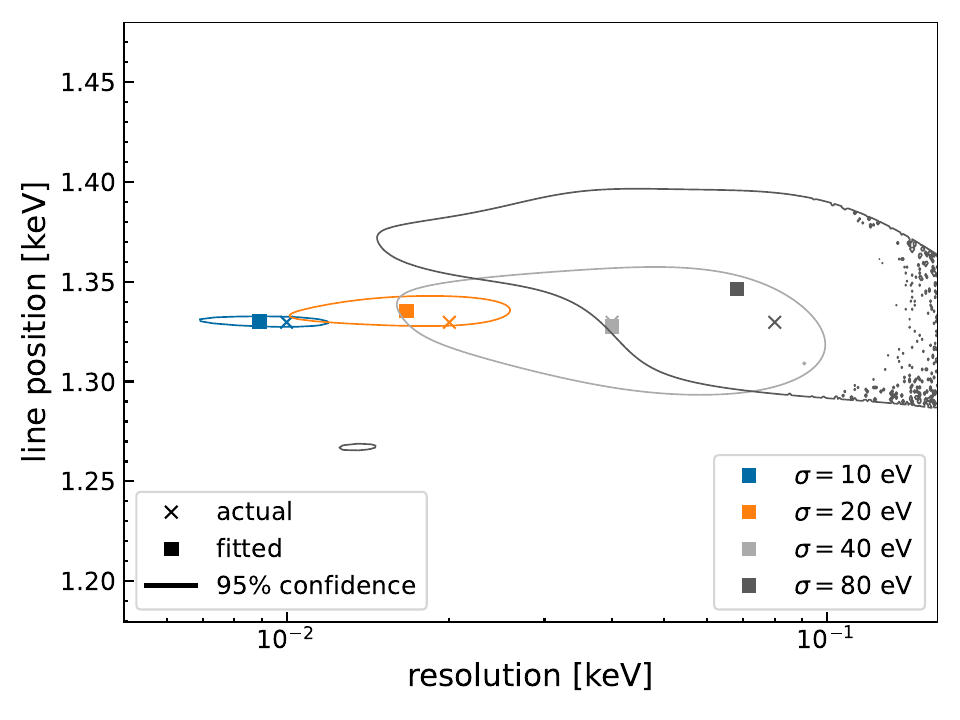}
    \caption{95\% confidence intervals (solid lines) and actual positions (X's) in peak resolution/position space for several four-week exposures at different detector resolutions. Squares show the fitted positions, the parameters that were found to maximize the likelihood over all values. The 20\,eV curves (orange) are for the sample shown in Figure~\ref{fig:sim-4wk-example}. Note extra regions in the 80\,eV contour occur because of fluctuations in the density of background events, causing local maxima in the likelihood function where the gaussian peak is fitted to these fluctuations. Widths of the confidence intervals are given in Table~\ref{tab:conf-spreads}.
    }
    \label{fig:sim-4wk-95contours}
\end{figure}

Figure~\ref{fig:sim-4wk-95contours} shows 95\% confidence intervals (for methods c.f.,~\cite{10.1103/PhysRevD.99.062001} \S VIIIC) for the extracted signal parameters \Efit{} and \sigfit{} for a few example four-week exposures at different detector resolutions, including the one pictured in Figure~\ref{fig:sim-4wk-example}. 
The horizontal and vertical spread of the contours
are given in Table~\ref{tab:conf-spreads}. For achievable resolutions, the signal peak position can be constrained to within 15\,eV -- for comparison, the expected shift in peak position due to Frenkel defect formation is on the order of 100\,eV (assuming linear scaling from about 15\,eV defect energy for 200\,eV NRs~\cite{PhysRevD.106.063012}).

\begin{table}[htbp]
    \centering
    \begin{ruledtabular}
    \begin{tabular}{ccc}
        Resolution [eV] & Width in \Efit{} [eV] & Width in \sigfit{} [eV] \\
        \hline
        10.0 & 5.296 & 5.027 \\
        \hline
        20.0 & 15.09 & 15.78 \\
        \hline
        40.0 & 63.98 & 83.2 \\
        \hline
        80.0 & 130.9 & 147.5 \\
    \end{tabular}
    \end{ruledtabular}
    \caption{Width of 95\% confidence intervals in \Efit{} (vertical direction) and \sigfit{} (horizontal direction) for the data shown in Figure~\ref{fig:sim-4wk-95contours}.}
    \label{tab:conf-spreads}
\end{table}

\section{\label{sec:conclusions}Discussion}

We have proposed a novel experiment to measure both products of thermal neutron capture on $^{29}$Si, the 8.47227\,MeV straight-to-ground postcapture gamma and the accompanying 1.32966\,keV nuclear recoil, in coincidence. Our planned experimental setup uses two 0.588\,kg Ge gamma detectors (\hmone{} and \hmtwo{}) to detect the deexcitation gamma and a one-gram Si phonon detector (\lm{}) to detect the NR. \texttt{Geant4} simulations show that with a 1\,mCi $^{252}$Cf source, we can achieve a positive measurement within a 28-day (26.1\,gram-day) exposure with existing detector technologies. This would be the first time such coincidence measurements are carried out for the products of neutron capture, and would be the first time the NR peak from neutron captures is directly measured in silicon. Furthermore, with existing technology we can measure the deposited NR energy with enough accuracy to detect the expected $\sim100\text{\,eV}$ shift due to defect formation.

The setup described here is the simplest, cheapest possible version of this experiment. Further modifications could be done to increase the efficiency of the method. First, the expected signal rate increases linearly with increasing phonon detector mass without substantially affecting the background in the \hmd{} detectors. With a 10\,g Si detector,  as long as there is no substantial loss in resolution, we could make a positive measurement in only 3\,days of exposure. Also, alternative neutron sources such as a D-D~generator or nuclear reactor would allow similar or much larger neutron fluence with a much lower level of correlated backgrounds, making detection easier. 

The current setup could also be modified to carry out specific measurements, such as a direct measurement of NR ionization yield with NTL-assisted phonon readout. 
In addition, the positioning of the \hmd{} detectors or the crystal orientation of the \lm{} detector could be modified to measure asymmetries in the postcapture recoil direction, and the \lm{} detector could be swapped out to measure neutron capture in different materials. 
Modifications to existing detector setups can also be made to facilitate \emph{in situ} detector calibration or yield studies. This proposed experiment is just one approach to making a measurement of this kind. This general method, to utilize specialized detectors to measure the twin products of neutron capture in coincidence, can be applied in myriad ways to study in detail the physics of nuclear recoils in matter.

\begin{acknowledgments}
ANV acknowledges support from the U.S. Department of Energy (DOE) Office of High Energy Physics
and from the National Science Foundation (NSF). ANV was supported by DOE Grant No. DE-SC0024275
and in part by NSF Grant No. 2111090. 
Portions of this research were conducted with the advanced computing resources provided by Texas A\&M High Performance Research Computing.
\end{acknowledgments}


\clearpage
\bibliography{capture.bib}
\bibliographystyle{apsrev4-2}

\end{document}